\newcommand\aj{{AJ\,}}%
\newcommand\araa{{ARA\&A\,}}%
\newcommand\apj{{ApJ\,}}%
\newcommand\apjl{{ApJ\,}}%
\newcommand\apjs{{ApJS\,}}%
\newcommand\aap{{A\&A\,}}%
\newcommand\mnras{{MNRAS\,}}%
\documentclass[graybox, natbib, footinfo]{svmult}

\usepackage{mathptmx}       
\usepackage{helvet}         
\usepackage{courier}        
\usepackage{type1cm}        
\usepackage{makeidx}         
\usepackage{graphicx}        
\usepackage{multicol}        
\usepackage[bottom]{footmisc}

\usepackage{mathrsfs}

\makeindex             

\begin{document}
\title*{The expected stellar populations in the {\it Kepler} and CoRoT fields}
\titlerunning{{\it Kepler}/CoRoT simulations} 
\author{L\'eo Girardi, Mauro Barbieri, Andrea Miglio, Diego Bossini, Alessandro Bressan, Paola Marigo, Tha\'{\i}se S. Rodrigues}
\authorrunning{Girardi et al.} 
\institute{L. Girardi, T. Rodrigues \at INAF -- Osservatorio Astronomico di Padova, Vicolo dell'Osservatorio 5, Padova, Italy \\ LIneA -- Laborat\'orio Interdepartamental de e-Astronomia, Rua Jos\'e Cristino, Rio de Janeiro, Brazil
\and M. Barbieri, P. Marigo \at Dipartimento di Fisica e Astronomia, Universit\`a di Padova, Italy 
\and A. Miglio, D. Bossini \at University of Birmingham, UK 
\and A. Bressan \at SISSA, Via Bonomea 265, I-34136 Trieste, Italy }
%
%
\maketitle

\abstract{
  Using the stellar population synthesis tool TRILEGAL, we discuss the expected stellar populations in the {\em Kepler} and CoRoT fields.}

\section{Inroduction}
\label{sec:intro}
{\it Kepler} and CoRoT asteroseismic observations are providing us
with precious information about the properties and structure of stars
displaced widely across the Galaxy. Observations of these fields will
not be repeated any time soon with instrumentation of comparable
precision and efficiency. Since both missions were primarily driven by
the goal of ``finding the most planets'', they applied complex target
selection criteria which are not ideal for the stellar populations and
Galaxy archeology applications which were devised later. Therefore,
what can be extracted from this data depends on a good understanding
of the data selection, and of our ability to model the entire samples
with population synthesis tools. This contribution will concentrate on
the latter aspect.

\section{Population synthesis of the Milky Way}
\label{sec:popsyn}

Population synthesis models of the Milky Way are the successors of the
``star counts'' models introduced in the 80's
\cite[e.g.][]{BahcallSoneira80, Bahcall86} to model the luminosity and
color distribution of stars across the sky. The main novelty
introduced by the population synthesis approach is the use of extended
databases of stellar models to describe the intrinsic luminosity
distribution of stars in given pass-bands, $\mathscr{L}_\lambda$,
instead of recurring to empirical data. More exactly,
$\mathscr{L}_\lambda$ are derived from grids of stellar evolutionary
tracks suitable converted into isochrones, which are then ``colored''
by using synthetic photometry applied to extended grids of model
spectra, and later weighted by assuming some star formation and
chemical enrichment history (SFH), and the initial mass function
(IMF). The all process is detailed in \cite[e.g.][]{Girardi_etal02,
  Girardi_etal05}. The advantages of the theoretical over the
empirically-derived $\mathscr{L}_\lambda$ are evident: there are
almost no limits to the kind of stellar populations to be tested;
moreover very different databases (e.g. comprising many passbands) can
be modeled in a consistent way, and the models can be more reliably
extrapolated to larger photometric depths. On the other hand, the
population synthesis approach introduces many additional parameters
and functions, like those describing the SFH and IMF for each Galactic
component, that apparently complicate the problem. Moreover, using the
population synthesis approach implies trusting on the predictive
capability of the underlying stellar models.

But the necessity of the population synthesis approach becomes
dramatically more evident when we consider the present asteroseismic
data: empirical data simply cannot replace the stellar models in this
case, simply because there is not enough empirical information to
build the asteroseismic versions of $\mathscr{L}_\lambda$ starting
from star clusters of from stars with parallaxes. In this case, the
asteroseismic data is helping to test the stellar models in a very
detailed way, star by star, and helping to test the galactic models at
the same time.

\subsection{TRILEGAL}
\label{subsec:tri}

TRILEGAL is a population synthesis code started with the initial goal
of simulating deep and wide multi-band photometric data
\citep{Groenewegen_etal02, Girardi_etal05, Vanhollebeke_etal09}. More
recently, the code has become a fundamental tool to test stellar
evolutionary tracks from the Padova-Trieste group
\citep[e.g.][]{Girardi_etal09, Girardi_etal10, Bianchi_etal11,
  Rosenfield_etal14}. With the recent/ongoing addition of new
quantities in the code, like the surface chemical composition and
asteroseismic parameters, it is ready to be applied in the simulation
and interpretation of large databases like those provided by CoRoT and
{\it Kepler}.

\subsection{Expectations for CoRoT}
\label{subsec:corot}

\citet{Miglio_etal13b, Miglio_etal13a} provide a detailed description
of the model expectations in CoRoT fields. CoRoT eyes are directed
towards very different lines-of sight, probing a large range of both
galactocentric radii and heights above/below the Galactic Plane. The
main result in \citet{Miglio_etal13a} was the detection of a
significant difference in the mass distribution of stars towards the
CoRoT fields LRc01 and LRa01, which was interpreted as being mainly
due to the different stellar ages being sampled at two different
heights below the Galactic Plane. Similar differences are expected for
all other CoRoT lines-of-sight, but are of harder interpretation given
their more complex target selection.

\subsection{Expectations for {\it Kepler}}
\label{subsec:kepler}

\begin{figure}[t]
\sidecaption[t]
\resizebox{0.99\hsize}{!}{\includegraphics{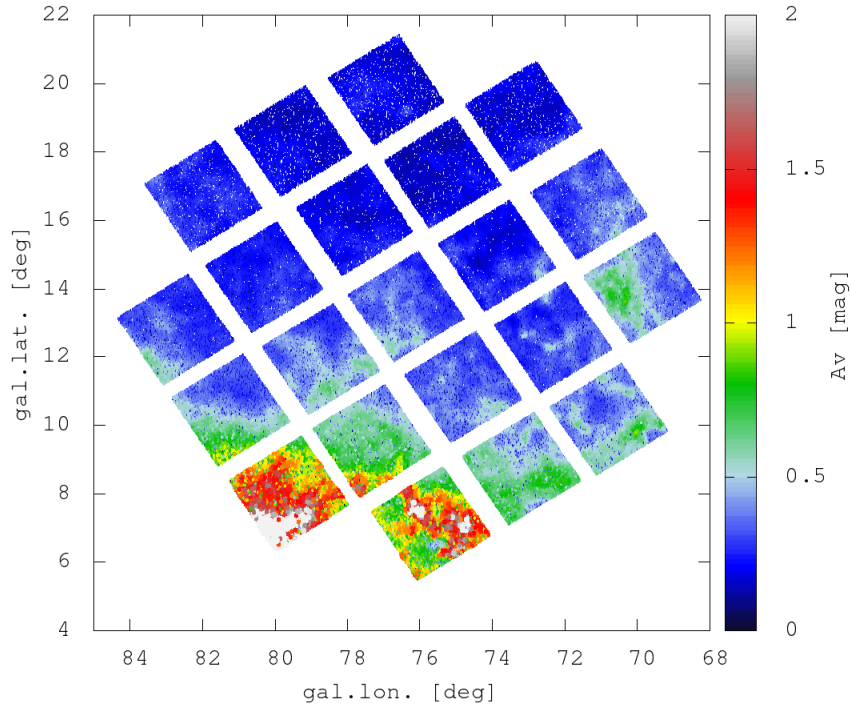}}
\caption{Barbieri et al.'s (in prep) large simulation of the {\it
    Kepler} field, including many of the filter systems of interest
  (e.g. SDSS, DOO51, 2MASS, and the wide $Kp$ filter). The plot simply
  shows the stellar location in the sky (galactic coordinates) color
  coded by their extinction $A_V$.}
\label{fig:keplerred}       
\end{figure}

Figure~\ref{fig:keplerred} is extracted from a large simulation of the
{\it Kepler} field (Barbieri et al. in prep), including many of the filter
systems of interest. The simulation scales down the
\citet{Schlegel_etal98} extinction values down to the distance of
each simulated star, distributing it along the line-of-sight by
assuming an exponential dust layer with $h_z=110$~pc. A correction for
the Local Bubble is also made.

One of the main novelties in the simulation are the more complex way
the binaries are simulated: their mass ratio, orbital period and
eccentricity, inclination etc., are derived for the system at birth
following a series of reasonable assumptions. The binary evolution is
then followed with the BSE code \citep{Hurley_etal02}, which considers
mass transfer and accretion, common-envelope evolution, collisions,
supernova kicks, angular momentum loss mechanisms, circularization and
synchronization of orbits by tidal interactions. Of course there are
many tunable parameters involved in these models, like the strength of
tidal damping in radiative, convective and degenerate regions, the
\cite{Reimers75} mass-loss coefficient, the binary enhanced mass
loss, the common envelope efficiency.
 

Any simulation of the {\it Kepler} field will be of limited use if not
including a simulation of the {\it Kepler} target selection criteria. The
best study of these criteria so far has been by \cite{Farmer_etal13},
who built a software that tries to mimic all the steps involved in
building the {\it Kepler} input catalogue and prioritization
\citep{Brown_etal11,Batalha_etal10}. For asteroseismic studies, we
should also consider the addition of targets not coming from the
original planet-detection plan, and estimate the probability of
actually measuring the asteroseismic parameters and their errors
\cite[e.g.][for dwarfs]{Chaplin_etal11}.

\begin{figure}[t]
\sidecaption[t]
\resizebox{0.99\hsize}{!}{\includegraphics{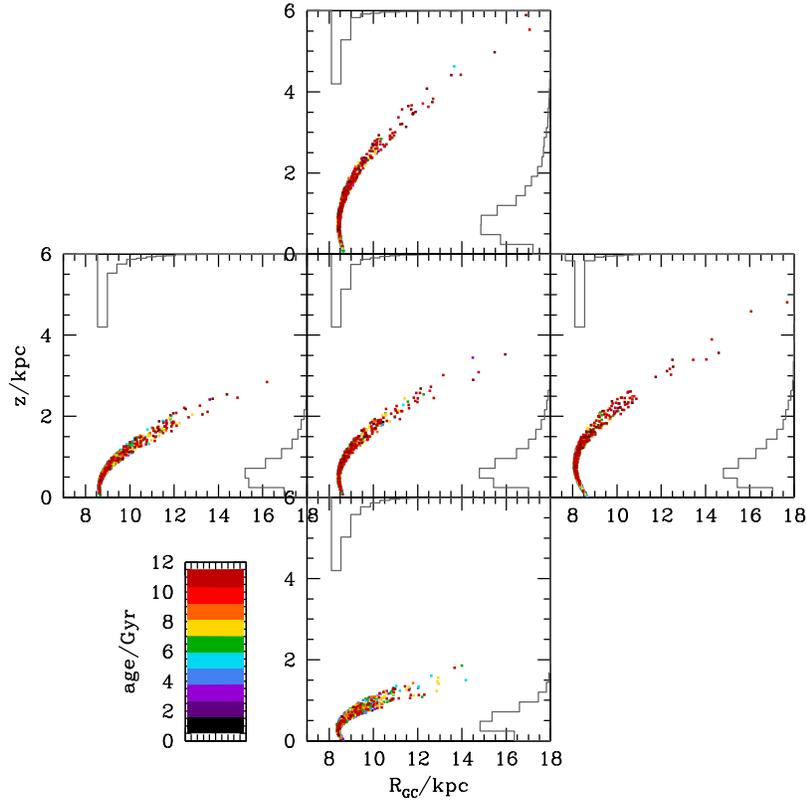}}
\caption{The expected ranges of galactocentric radius, $R_{\rm GC}$,
  and height above the Plane, $z$, for subsamples located in the N, S,
  W and E extreme corners of the {\it Kepler} field, and for the center as
  well. It evidences the modest coverage in $R_{\rm GC}$, and the
  large range in $z$. Younger stars are concentrated at smaller
  heights.}
\label{fig:keplerred2}       
\end{figure}

\begin{figure}[t]
\sidecaption[t]
\resizebox{0.49\hsize}{!}{\includegraphics{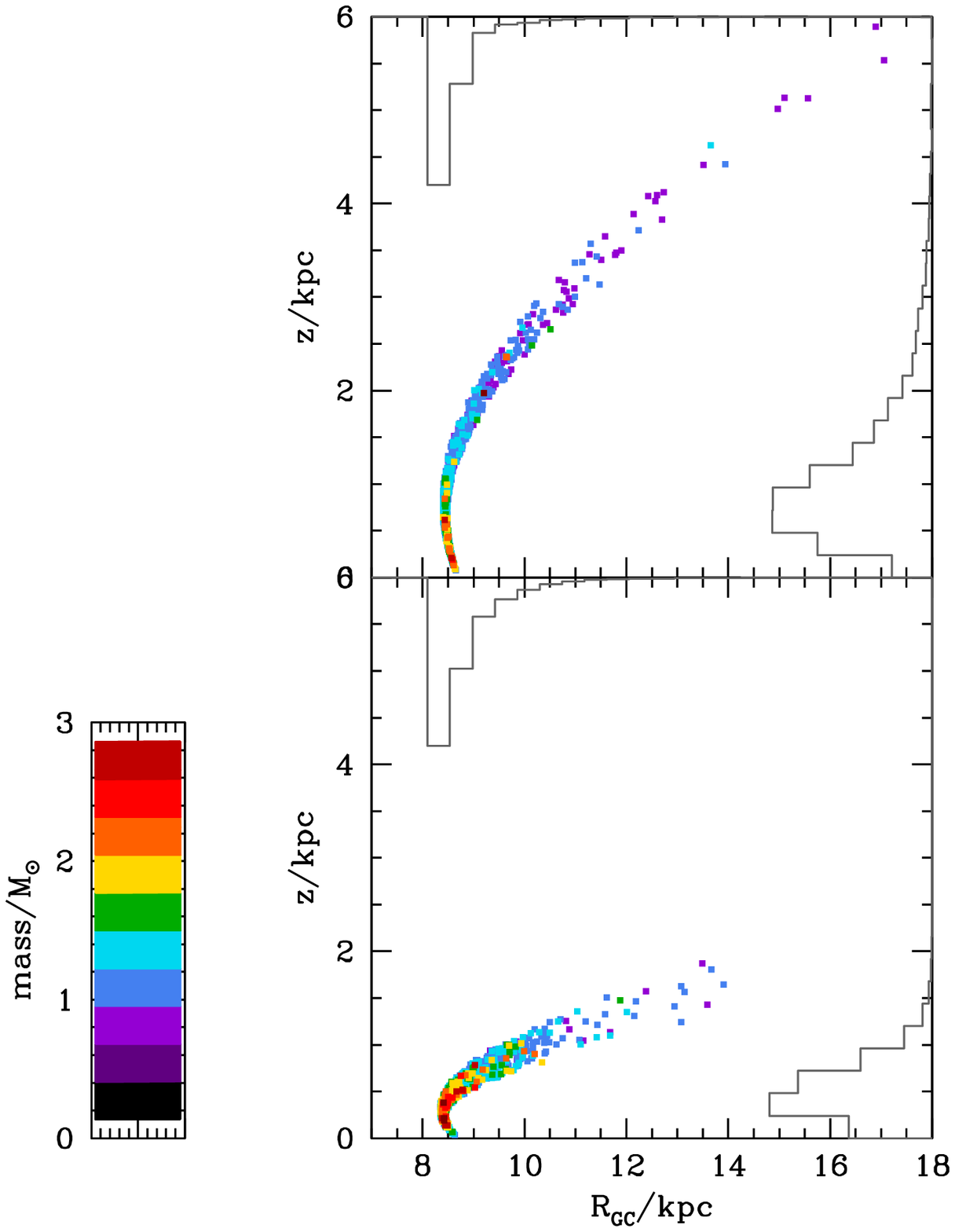}}
\resizebox{0.49\hsize}{!}{\includegraphics{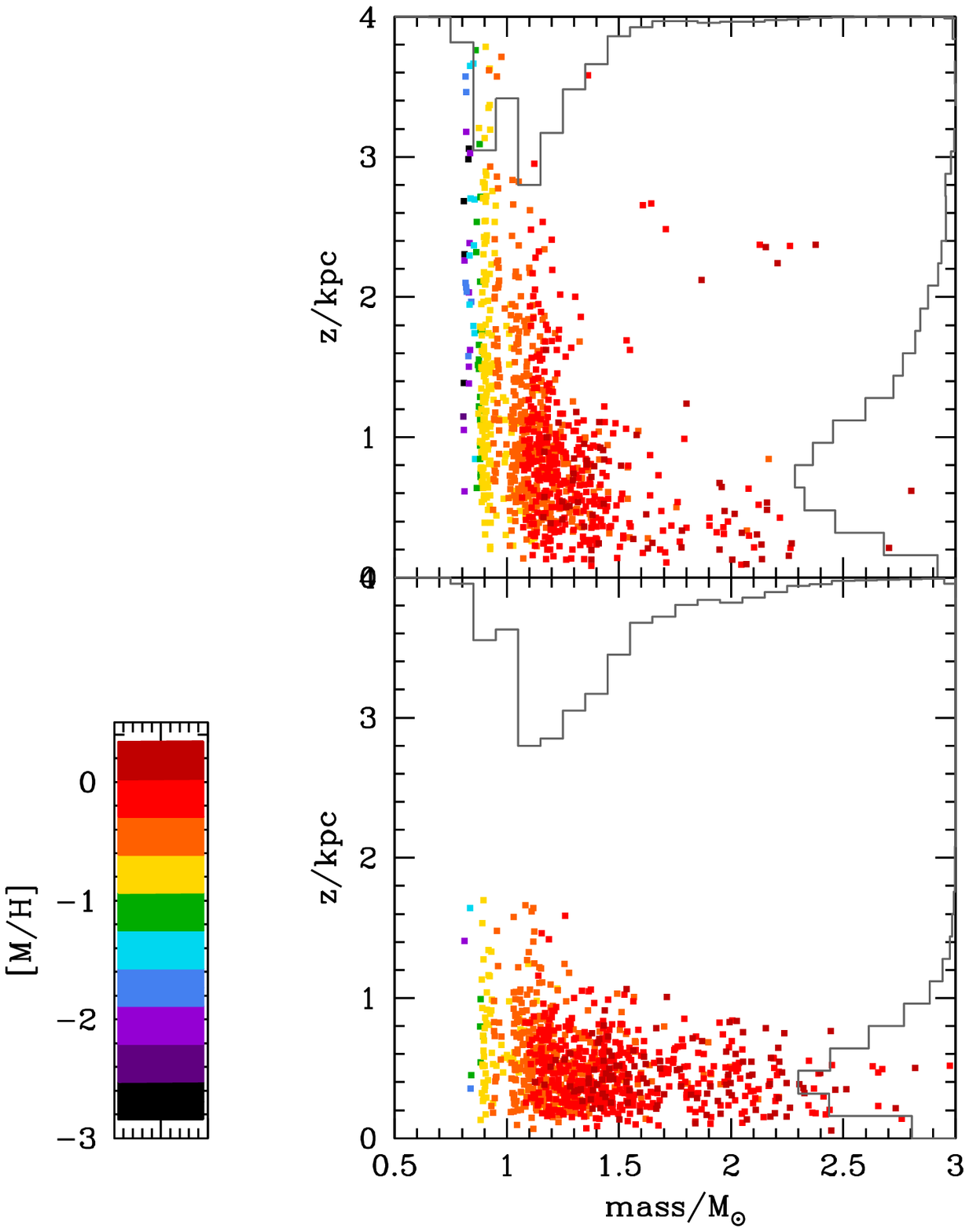}}
\caption{Left panel: distribution of heights above the plane for the
  two extreme fields of {\it Kepler} (top for $b=20^\circ$, bottom for
  $b=6^\circ$). Right panel: The expected distribution of stellar
  masses for stars in these fields.}
\label{fig:keplerred3}       
\end{figure}

Figures~\ref{fig:keplerred2} provides some basic information about the
expected distribution of these stars across the Galaxy, in particular
evincing the modest coverage in $R_{\rm GC}$, and the large range in
$z$. Since younger stars are concentrated at smaller heights, the
model predicts strong differences between the distributions stellar
masses observed at the latitude extremes of this field, as evidenced
in \ref{fig:keplerred3}. These differences are large enough to be
easily measurable in the {\it Kepler} data, although their interpretation is
somewhat complicated by the patchy extinction in the low-latitude
fields. Work is ongoing to transform these distributions of stellar
mass in clear constraints to the increase of scale-height with stellar
age, hence complementing the earlier suggestions derived from CoRoT
data.

\begin{figure}[t]
\sidecaption[t]
\resizebox{0.49\hsize}{!}{\includegraphics{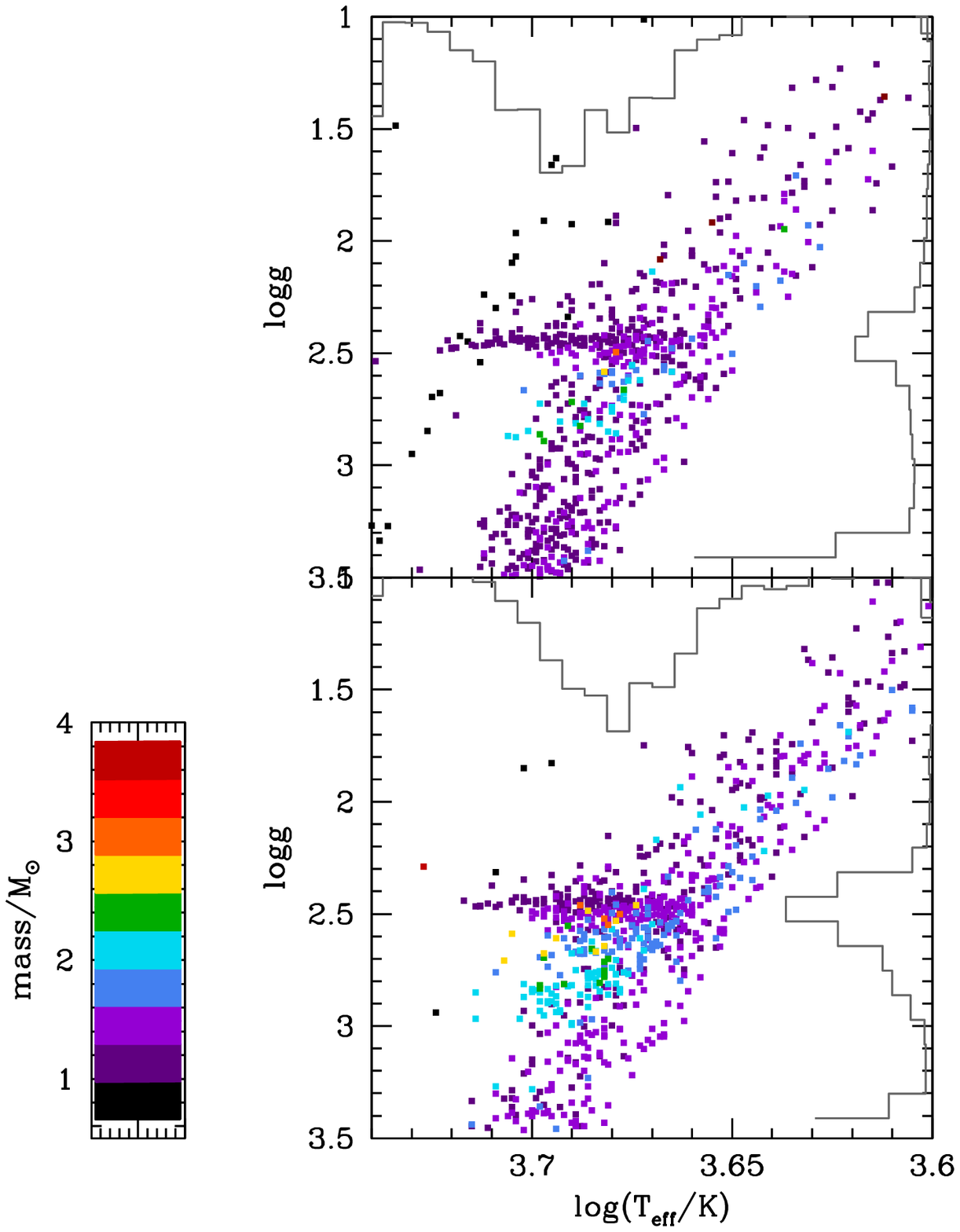}}
\resizebox{0.49\hsize}{!}{\includegraphics{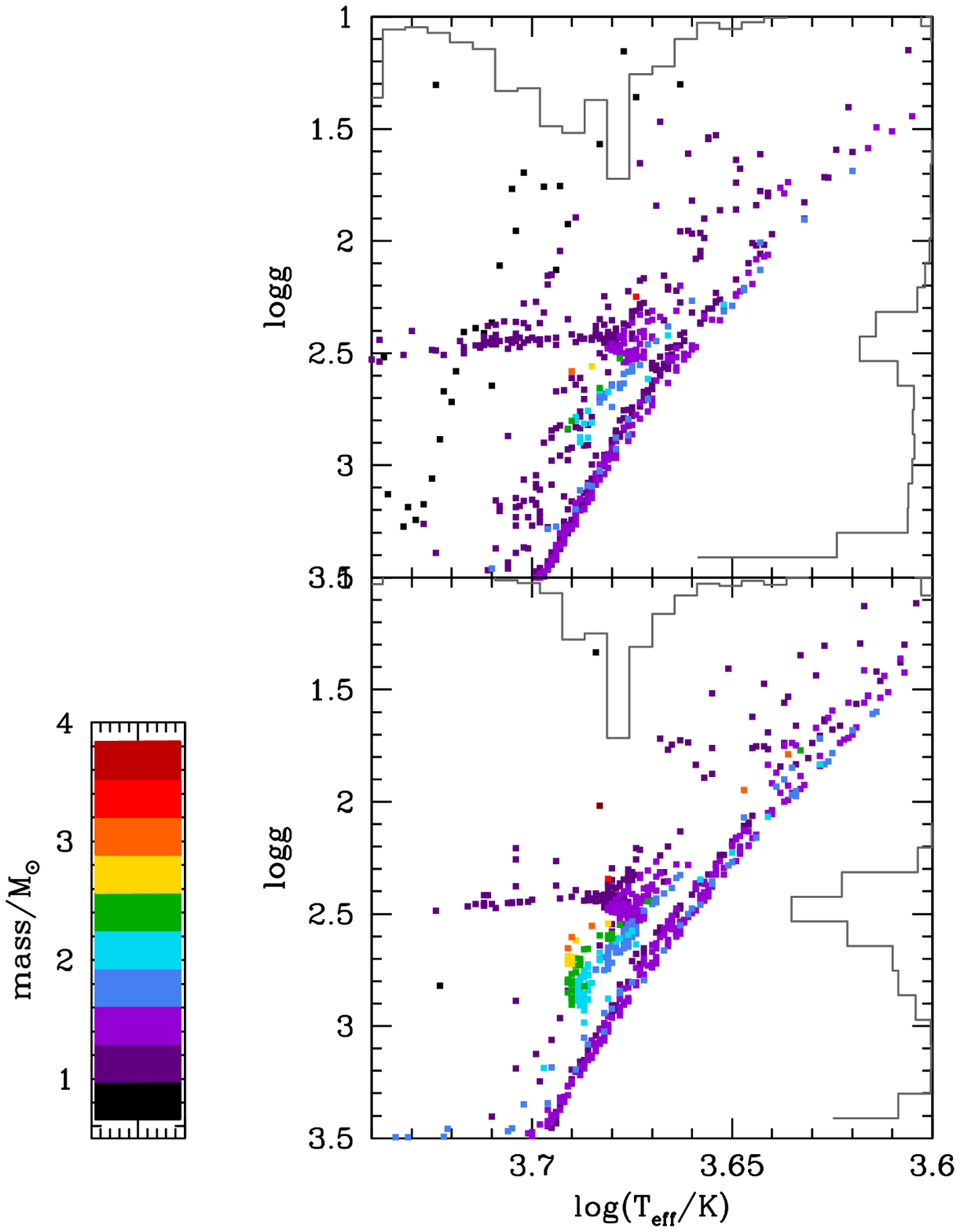}}
\caption{Left panel: red clump with default metallicity
  dispersion. Right panel: without metallicity dispersion}
\label{fig:rc}       
\end{figure}

It is hoped that {\it Kepler} data will help to solve the dramatic problem
pointed out by \cite{ReyleRobin01}, that is: when attempting to fit
star counts only, there is a strong degeneracy between scale heigth
and surface density of thick disk, with models favoring either high
scale height and small local density (for example, the $h_z=1400$~pc
and 2\% of disk density favoured by \citealt{ReidMajewski93}, against the
910~pc and 5.9\% from \citealt{Buser_etal99}). Although the problem can
be relieved with more photometric data, a clearcut distinction between
thick and from old thin disk (if any exists,  e.g. \citet{Bovy_etal12})
would help. That is where {\it Kepler} data can be critical.

Additional information is store in {\it Kepler} data in the form of
chemical abundances and kinematics of stars of different masses and
ages. The secondary red clump \citep{Girardi99}, for instance,
contains a pure population of $\sim$1~Gyr stars and which is easily
identifiable in a $\log g$ versus $\log T_{\rm eff}$ plot
(Fig.~\ref{fig:rc}), especially when additional information from mixed
modes (period spacing) is available. Stellar metallicities for these
stars, as measured e.g. by APOKASC \citep{APOKASC} can provide a
direct probe of the intrinsic metallicity spread, and a solid point
along the age--metallicity relation, across the Galaxy.

\section{Concluding remarks}
\label{sec:conclu}
The population synthesis approach has revealed to be a powerful
technique, very useful for the interpretation of wide-area surveys in
terms of the MW structure and evolution. Present applications to CoRoT
and {\it Kepler} asteroseismic samples are still limited, but with their
direct measurements of ages and evolutionary stages, they provide
excellent hopes for imposing tight constrains in the models.

While dwarfs in {\it Kepler} fields seem to have their radii well reproduced
by models, there is a discrepancy for masses, still to be clarified
\citep{Chaplin_etal11}. Giants in {\it Kepler} represent the ideal sample for
testing the variation of stellar properties with $z$, and hopefully
will provide long-awaited constraints to disk-heating and accretion
scenarios, and complement the kinematical information either already
available (e.g. from proper motions, GCS) or being collected by
APOKASC. The expected variation of mean mass (age) with $z$ is
detected in the CoRoT giants \citep{Miglio_etal13a}, but those results
may be affected by the large range of galactocentric radii probed by
CoRoT. Of course, it is expected that once the vertical structure is
revealed by the {\it Kepler} sample, CoRoT results will have to be
reevaluated.

\begin{acknowledgement}
We thank all participants in the Sexten meeting for the stimulating
discussions, and the {\it Kepler} and CoRoT teams for providing such
wonderful and promising databases.
\end{acknowledgement}

%


\end{document}